\documentclass[twocolumn]{aastex701}

\usepackage{graphicx}
\usepackage{subcaption}
\usepackage{multirow}
\usepackage{array}
\usepackage{rotating} 
\usepackage{booktabs}
\usepackage{tikz}
\usepackage{hyperref}
\usepackage{threeparttable}

\begin{document}

\title{Short Spike, Long Story: Episode-Dependent Shifts of Long-Duration Type-I GRBs on $E_{\rm p,z}$--$E_{\rm iso}$ Plane}

\author[0009-0004-7110-3450]{Ya-Hui Jiang}
\affiliation{School of Astronomy and Space Science, Nanjing University, Nanjing 210093, People's Republic of China}
\affiliation{Key Laboratory of Modern Astronomy and Astrophysics (Nanjing University), Ministry of Education, People's Republic of China}
\email{yahjiang@smail.nju.edu.cn}

\author[0009-0009-2083-1999]{Run-Chao Chen}
\affiliation{School of Astronomy and Space Science, Nanjing University, Nanjing 210093, People's Republic of China}
\affiliation{Key Laboratory of Modern Astronomy and Astrophysics (Nanjing University), Ministry of Education, People's Republic of China}
\email{chrczxx@smail.nju.edu.cn}

\author[0009-0008-2841-3065]{Zhen-Yu Yan}
\affiliation{School of Astronomy and Space Science, Nanjing University, Nanjing 210093, People's Republic of China}
\affiliation{Key Laboratory of Modern Astronomy and Astrophysics (Nanjing University), Ministry of Education, People's Republic of China}
\email{xxx@smail.nju.edu.cn}

\author[0009-0005-1517-6089]{Kamil Nadaf}
\affiliation{School of Astronomy and Space Science, Nanjing University, Nanjing 210093, People's Republic of China}
\affiliation{Key Laboratory of Modern Astronomy and Astrophysics (Nanjing University), Ministry of Education, People's Republic of China}
\email{xxx@smail.nju.edu.cn}

\author[0000-0003-3659-4800]{Xiao-Hong Zhao}
\affiliation{Yunnan Observatories, Chinese Academy of Sciences, Kunming, People’s Republic of China}
\affiliation{Center for Astronomical Mega-Science, Chinese Academy of Sciences, Beijing, People’s Republic of China}
\email{xxx@smail.nju.edu.cn}
\author[0000-0003-4111-5958]{Bin-Bin Zhang}
\affiliation{School of Astronomy and Space Science, Nanjing University, Nanjing 210093, People's Republic of China}
\affiliation{Key Laboratory of Modern Astronomy and Astrophysics (Nanjing University), Ministry of Education, People's Republic of China}
\email[show]{bbzhang@nju.edu.cn}

\begin{abstract}

Observations of GRB have increasingly challenged the traditional $T_{90}$-based classification, demonstrating that duration does not map uniquely onto progenitor type. A particularly striking class consists of long-duration Type~I GRBs (merger-origin) events whose prompt emission lasts far longer than the 2 s boundary, typically comprising an initial short hard spike followed by softer extended emission. Identifying the physical origin of such bursts requires diagnostics beyond duration alone, among which the $E_{\rm p,z}$--$E_{\rm iso}$ relation is widely used as a complementary classification tool. We analyze a sample of eight long-duration Type~I GRBs and merger candidates by separating the initial spike from the subsequent extended emission and examining their episode-dependent locations on the $E_{\rm p,z}$--$E_{\rm iso}$ plane. We find that the initial spike generally lies within, or close to, the empirical Type~I region, consistent with a compact-merger-like prompt-emission component. In contrast, the extended-emission episode systematically occupies a region closer to that of Type~II GRBs, and could be misidentified as collapsar-like if analyzed in isolation. This Type~I-to-Type~II transition is further supported by time-resolved spectral analysis, suggesting diversity in central-engine evolution or outflow properties between the two emission phases. Our results caution that the Amati relation alone can lead to a misleading empirical classification when the initial hard spike is weak, falls outside the instrumental bandpass, or is missed entirely, leaving only the extended emission to be analyzed. Broad temporal and spectral coverage, together with independent multi-wavelength diagnostics, is therefore essential for identifying the physical origin of these complex events.

\end{abstract}

\keywords{\uat{Gamma-ray bursts}{629}}

\section{Introduction} 

Gamma-ray bursts (GRBs) have traditionally been divided into two populations according to their prompt-emission duration, with a boundary at $T_{90}=2~\mathrm{s}$ separating short and long GRBs \citep{1993ApJ...413L.101K, 2013ApJ...764..179B}. The physical significance of this phenomenological division became apparent only later through the discovery of distinct multi-messenger counterparts. Long GRBs were found to be associated with broad-lined Type Ic supernovae \citep{1998Natur.395..670G, 2006ARA&A..44..507W}, while short GRBs were linked to compact-object mergers \citep{2006ApJ...648.1110B, 2007PhR...442..166N} through the detection of r-process nucleosynthesis signatures \citep{1974ApJ...192L.145L, 1999ApJ...525L.121F} in their afterglow spectra (i.e., kilonova or mergernova; \citealt{1998ApJ...507L..59L, 2010MNRAS.406.2650M}), and, in some cases, corroborated by gravitational-wave observations \citep{2017PhRvL.119p1101A}. As a result, GRB duration came to be widely regarded as a practical observational proxy for progenitor type, with long and short GRBs broadly corresponding to collapsar-origin and merger-origin events, respectively.

Yet this simple duration--progenitor mapping has been challenged by notable exceptions on both sides. GRB~060614 was among the first events to question the traditional paradigm: it exhibited long duration and extended soft emission, yet no accompanying supernova was detected down to deep limits at its relatively low redshift \citep{2006Natur.444.1050D, 2006Natur.444.1044G}. Years later, GRB~211211A provided far stronger evidence that long-duration GRBs can arise from compact-object mergers, as its extended high-energy emission was accompanied by a clearly detected kilonova counterpart \citep{2022Natur.612..232Y, 2022Natur.612..223R}. More recently, GRB~230307A further demonstrated that merger-origin GRBs can produce extremely energetic and temporally complex prompt emission otherwise characteristic of collapsars \citep{2025NSRev..12E.401S, 2024Natur.626..742Y}. The breakdown runs in the opposite direction as well: GRB~200826A appeared as a short-duration event in its rest frame, yet subsequent multi-wavelength observations revealed compelling evidence for a collapsar origin, including a clear supernova association \citep{2021NatAs...5..911Z, 2021NatAs...5..917A, 2022ApJ...932....1R}. Taken together, these discoveries demonstrate that duration alone does not unambiguously reveal the physical origin of a GRB, and that the correspondence between $T_{90}$ and progenitor type is far less robust than previously assumed.

A variety of supplementary diagnostics have therefore been proposed to infer GRB progenitors, including supernova or kilonova associations, host-galaxy properties, multi-wavelength afterglow behavior, and empirical prompt-emission correlations. Among them, one of the most widely used is the \textit{Amati} relation \citep{2002A&A...390...81A}, which links the rest-frame spectral peak energy $E_{\rm peak}$ and the isotropic-equivalent radiated energy $E_{\rm iso}$. The relation was originally established using long GRBs with measured redshifts, which form a well-defined, approximately linear sequence in the $\log E_{\rm peak}$--$\log E_{\rm iso}$ plane \citep{2002A&A...390...81A, 2006MNRAS.372..233A, 2017ApJ...850..161T}. As the number of short GRBs with secure redshift determinations gradually increased, these events were found to occupy a systematically different region of the same parameter space, exhibiting higher $E_{\rm peak}$ and lower $E_{\rm iso}$ than typical long GRBs \citep{2012ApJ...750...88Z, 2018NatAs...2...69Z, 2015MNRAS.451..126S}. The two populations broadly occupy two distinct regions of the $E_{\rm p,z}$--$E_{\rm iso}$ plane, commonly referred to as Type~I and Type~II, corresponding approximately to merger-origin and collapsar-origin GRBs, respectively \citep{2009ApJ...703.1696Z,2010ApJ...725.1965L, 2020MNRAS.492.1919M}. The Amati relation has therefore become an important complementary classification tool for peculiar GRBs whose physical origins cannot be unambiguously inferred from duration alone.

However, recent observations of multi-episode GRBs suggest that a burst's location on the $E_{\rm p,z}$--$E_{\rm iso}$ plane may be more ambiguous than commonly assumed, as different emission episodes within the same event can exhibit markedly different spectral properties. GRB~160425A, for instance, contains an initial short-hard spike and a subsequent long-duration emission component, with the two episodes falling in opposite regions of the $E_{\rm p,z}$--$E_{\rm iso}$ plane \citep{2026arXiv260328699L}. A related complication arises from incomplete spectral coverage: EP250704a/GRB~250704B exhibits a short hard gamma-ray spike consistent with a Type~I classification, while its subsequent long-lasting soft X-ray emission --- only revealed by the broad soft-X-ray coverage of the Einstein Probe mission \citep{2022hxga.book...86Y} --- aligns more closely with Type~II expectations \citep{2026arXiv260114137L}. This illustrates that a significant fraction of GRB emission may remain hidden outside the conventional gamma-ray band, and that the inferred Amati-relation-based classification of a burst can change substantially depending on which emission components happen to be detected. 
These examples suggest that Amati-relation based classification can depend on which emission episode is included in the analysis, especially when different episodes occupy substantially different spectral regimes. This raises a direct question: if the initial spike and the extended emission are analyzed separately, do they occupy the same region of the $E_{\rm p,z}$--$E_{\rm iso}$ plane, or can the inferred empirical classification change from one episode to the other?

Motivated by this question, we systematically investigate a sample of eight long-duration Type~I GRBs and merger candidates --- events whose prompt emission substantially exceeds the canonical 2~s boundary yet shows observational indications of a compact-merger origin, each exhibiting an initial short hard spike followed by softer extended emission. By separating these two components and examining their individual locations on the $E_{\rm p,z}$--$E_{\rm iso}$ plane, we assess the robustness of Amati-relation based progenitor classification for this population and characterize how episode-dependent spectral evolution drives the apparent transition between the two empirical regions.

The remainder of this paper is organized as follows. Section~\ref{sec:data} describes the sample selection, data reduction, and spectral analysis methods. Section~\ref{sec:results} presents the temporal and spectral properties of the sample, focusing on the episode-dependent behavior on the $E_{\rm p,z}$--$E_{\rm iso}$ plane. Section~\ref{sec:discussion} discusses the implications of our results, including misclassification risks, the diversity of evolutionary trends, and possible physical interpretations. Section~\ref{sec:conclusion} summarizes our main conclusions.

\section{Sample Selection and Data Analysis} \label{sec:data}

\subsection{Sample Selection and Data Reduction}

Our study focuses on long-duration Type~I GRBs and Type~I candidates: events whose total prompt-emission durations exceed the canonical $T_{90}=2$~s boundary, but which are either known or suspected to originate from compact-object mergers. To construct the sample, we selected GRBs satisfying at least one of the following criteria: (1) long-duration GRBs with direct observational indications of a merger origin, such as an associated kilonova or the absence of a supernova at low redshift; or (2) long-duration, multi-episode GRBs exhibiting the characteristic Type~I-like prompt-emission morphology of an initial short hard spike followed by a longer-duration softer emission component. The first criterion selects confirmed or strong merger candidates, while the second selects morphological analogs whose progenitor type is not yet independently established; the latter group allows us to test whether the episode-dependent behavior identified in confirmed mergers extends to the broader population sharing the same prompt-emission structure.

The final sample is summarized in Table~\ref{tab:sample}. GRB~250704B is included as a special case, as its extended emission component is detected exclusively in the soft X-ray band. For this burst we adopt the light-curve and spectral results directly from \citet{2026arXiv260114137L}. Three episodes are reported in that work, of which we take only Episode~I and Episode~III for our analysis, corresponding to the spike and extended-emission components, respectively. Episode~II, a faint signal separated from Episode~I by a clear temporal gap, is excluded, consistent with our general criterion of discarding faint intermediate signals between the spike and the extended emission. For each burst, Table~\ref{tab:sample} lists the redshift, instruments used, and the key observational properties motivating its inclusion --- specifically, evidence for a merger origin and/or the presence of a distinct short-hard spike followed by extended emission. We emphasize that bursts selected solely by criterion (2), such as GRB~100212A, should be regarded as long-duration Type~I \textit{candidates}; their inclusion serves to test the generality of the episode-dependent behavior rather than to presume their progenitor type.

Since emission in these bursts commonly evolves from hard gamma rays to softer X-ray bands, multi-band coverage is essential for capturing both emission episodes. Our analysis therefore primarily relies on Swift/BAT \citep{2004ApJ...611.1005G, 2013ApJS..209...14K} as the core instrument, supplemented by Fermi/GBM \citep{2009ApJ...702..791M} and Swift/XRT \citep{2005SSRv..120..165B} data when available.

The Swift/BAT data were processed using the \texttt{BatAnalysis} package \citep{batanalysis2025,2025ApJ...988...23P,2023HEAD...2010301P}, a Python-based analysis tool built upon the \texttt{HEASoft} software suite\footnote[1]{\url{https://heasarc.gsfc.nasa.gov/docs/software/heasoft/}}. Standard mask-weighted light-curve extraction, spectral extraction, and response generation were applied to all bursts in the sample.

To extend the spectral coverage toward higher energies, we incorporated Fermi/GBM observations when available. For each burst, we selected up to two sodium iodide (NaI) detectors with source incident angles smaller than $60^\circ$, prioritizing those with the smallest source angles. The bismuth germanate (BGO) detector with the smallest source incident angle was also included to further extend the high-energy coverage. The GBM data were processed using the \texttt{Heapy}\footnote[2]{\url{https://github.com/jyangch/heapy/}} package, and the spectral slices were selected to match those defined from the BAT analysis.

For bursts with soft X-ray observations, Swift/XRT data were further included to supplement the low-energy spectral coverage. Standard XRT reduction procedures \citep{2007A&A...469..379E, 2009MNRAS.397.1177E}, including pile-up correction, light-curve extraction, and spectral extraction, were performed using \texttt{HEASoft}. 

For GRB~230307A, significant pulse pile-up, deadtime, and telemetry saturation in the GBM data render the brightest emission intervals unreliable; we therefore adopted GECAM-B detector-4 observations instead, following the reduction procedures described in \cite{2025NSRev..12E.401S}.

\begin{table*}[htbp]
 \centering
 \caption{Information of the GRB sample adopted in this work.}
 \label{tab:sample}
 \begin{tabular}{c c c l l}
 \toprule
 GRB name & Redshift & Instrument & Inclusion Reason & Episode ranges (s) \\
 \midrule
 \multirow{2}{*}{060614} & \multirow{2}{*}{0.125$^{\textsuperscript{a}}$} & \multirow{2}{*}{Swift} & \multirow{2}{*}{Suggestive kilonova association$^{\textsuperscript{g}}$} & I:~~$[-1.55,\ 3.25]$ \\
 & & & & II:~$[4.34,\ 99.95]$ \\[4pt]
 \multirow{2}{*}{080503} & \multirow{2}{*}{Unknown} & \multirow{2}{*}{Swift} & \multirow{2}{*}{Suggestive kilonova association$^{\textsuperscript{h}}$} & I:~~$[0.11,\ 0.62]$ \\
 & & & & II:~$[17.07,\ 93.04]$ \\[4pt]
 \multirow{2}{*}{100212A} & \multirow{2}{*}{Unknown} & \multirow{2}{*}{Swift, Fermi} & \multirow{2}{*}{Short spike + long-duration emission} & I:~~$[-0.38,\ 1.02]$ \\
 & & & & II:~$[75.33,\ 81.98]$ \\[4pt]
 \multirow{2}{*}{160425A} & \multirow{2}{*}{0.555$^{\textsuperscript{b}}$} & \multirow{2}{*}{Swift} & \multirow{2}{*}{Short spike + long-duration emission$^{\textsuperscript{i}}$} & I:~~$[-0.54,\ 1.38]$ \\
 & & & & II:~$[260.96,\ 289.12]$ \\[4pt]
 \multirow{2}{*}{211211A} & \multirow{2}{*}{0.076$^{\textsuperscript{c}}$} & \multirow{2}{*}{Swift, Fermi} & \multirow{2}{*}{Kilonova association$^{\textsuperscript{j}}$} & I:~~$[0.03,\ 11.49]$ \\
 & & & & II:~$[12.06,\ 53.47]$ \\[4pt]
 \multirow{2}{*}{211227A} & \multirow{2}{*}{0.228$^{\textsuperscript{d}}$} & \multirow{2}{*}{Swift} & \multirow{2}{*}{Suggestive kilonova association$^{\textsuperscript{k}}$} & I:~~$[0.03,\ 1.82]$ \\
 & & & & II:~$[10.72,\ 68.19]$ \\[4pt]
 \multirow{2}{*}{230307A} & \multirow{2}{*}{0.065$^{\textsuperscript{e}}$} & \multirow{2}{*}{Fermi, GECAM} & \multirow{2}{*}{Kilonova association$^{\textsuperscript{l}}$} & I:~~$[-0.2,\ 0.2]$ \\
 & & & & II:~$[0.4,\ 48]$ \\[4pt]
 \multirow{2}{*}{250704B} & \multirow{2}{*}{0.661$^{\textsuperscript{f}}$} & \multirow{2}{*}{SVOM, EP} & \multirow{2}{*}{Short spike + long-duration emission$^{\textsuperscript{m}}$} & I:~~$[-0.1,\ 0.4]$ \\
 & & & & II:~$[22.76,\ 764.50]$ \\
 \bottomrule
 \end{tabular}
 \begin{tablenotes}
 \small
 \item \textbf{References:}
 ($a$) \citet{2006GCN..5275....1P};
 ($b$) \citet{2016GCN.19350....1T};
 ($c$) \citet{2021GCN.31221....1M};
 ($d$) \citet{2021GCN.31324....1M};
 ($e$) \citet{2023GCN.33485....1G};
 ($f$) \citet{2025GCN.40966....1A};
 ($g$) \citet{2015NatCo...6.7323Y};
 ($h$) \citet{2015ApJ...807..163G};
 ($i$) \citet{2026arXiv260328699L};
 ($j$) \citet{2022Natur.612..232Y};
 ($k$) \citet{2022ApJ...931L..23L};
 ($l$) \citet{2025NSRev..12E.401S};
 ($m$) \citet{2026arXiv260114137L}.
 \end{tablenotes}
\end{table*}

\subsection{Definition of Temporal Episodes}\label{subsec:seg}

For each burst, we divide the prompt emission into two components: an initial short-duration spike (Episode~I) and a subsequent longer-duration emission component (Episode~II). The temporal boundaries of the two episodes are listed in Table~\ref{tab:sample}.

The traditional $T_{90}$ characterizes the overall duration of a burst as a whole, but does not provide the boundaries between the initial spike and the subsequent extended emission that our episode-based analysis requires. We therefore identify different emission episodes by applying the Bayesian Blocks algorithm \citep{2013ApJ...764..167S} to the mask-weighted Swift/BAT light curves\footnote[3]{GRB~230307A is an exception, for which we adopt the episode definition based on the Fermi/GBM light curve.} for the following two reasons.  First, the mask-weighted BAT light curves become strongly affected by statistical fluctuations when the SNR is low, making duration estimates based on cumulative counts less reliable for weak extended emission \citep{2010ApJ...717..411N, 2022ApJ...927..157M}. Second, several bursts in our sample exhibit continuous emission above the background without a clear quiescent gap between the initial spike and the later emission, so that the two components cannot be separated on the basis of a single global duration measurement. The Bayesian Blocks algorithm instead adaptively identifies statistically significant temporal structures while suppressing background fluctuations. Episode boundaries were then defined by grouping contiguous blocks according to the overall light-curve morphology, using either the presence of a quiescent interval or a significant change in count-rate level as the dividing criterion.

We note that Episode~I formally exceeds 2~s for GRB~060614 and GRB~211211A. We nevertheless retain this definition for both events, as they possess independent observational evidence supporting a merger origin \citep{2015NatCo...6.7323Y,2022Natur.612..232Y}, and their light curves still exhibit a clear morphological separation between an initial spike-like component and a later extended component \citep{2007ApJ...655L..25Z,2022Natur.612..223R}. In these cases, the episode definition is therefore driven by light-curve morphology rather than the $T_{90}=2$~s boundary.

GRB~230307A represents a more complex case. Its prompt emission consists of a bright short spike followed by a prominent FRED-like bump. The latter component itself contains multiple energy-dependent sub-pulses and an apparently achromatic temporal dip at $\sim$18~s after the trigger \citep{2025NSRev..12E.401S,2024ApJ...969...26P, 2025ApJ...979...73W}, making a clean episode boundary difficult to define from temporal structure alone. We therefore adopt a morphology-based definition and assign the entire FRED-like bump to Episode~II.

The time ranges of Episodes~I and II for all bursts in our sample, determined following the procedures described above, are listed in Table~\ref{tab:sample}.

\subsection{Spectral Analysis}

Spectral fitting was performed using the \texttt{BaySpec}\footnote[4]{\url{https://github.com/jyangch/bayspec/}} package. For bursts with multi-instrument observations, joint spectral fitting was conducted using the corresponding BAT, GBM, and XRT data within the same temporal intervals listed in Table~\ref{spectable}. We considered several spectral models: the power-law (PL), cutoff power-law (CPL), blackbody (BB), and, when data quality permitted, the Band function \citep{1993ApJ...413..281B}. Generally, the choice of the statistic for model selection depends on the ratio between the number of data points and free parameters \citep{burnham2002model}. In our sample, however, this condition is not uniform: GECAM spectra and joint fits with multiple instruments have \(N_\mathrm{data}\) ranging from 90 to 350, whereas BAT-only spectra typically have \(N_\mathrm{data}\) of several tens. As a result, the \(N_\mathrm{data}/N_\mathrm{param}\) ratio varies across our sample, with some fits satisfying \(N_\mathrm{data}/N_\mathrm{param}>40\) and others not. To maintain a consistent standard across the whole sample, we adopted the Bayesian Information Criterion \citep[BIC;][]{1978AnSta...6..461S} values as our primary criterion. For completeness, we also compute the corrected AIC \textbf{\citep[AICc;][]{burnham2002model}} as a cross‑check. The two criteria yield consistent results for nearly all episodes; only GRB 211227A Episode~II differs, and this discrepancy does not affect our main conclusions.

The peak energy $E_{\rm peak}$ was derived from the best-fit model as follows. For the CPL and Band models, $E_{\rm peak}$ was directly obtained from the fitted spectral parameters. For the BB model, we adopted $E_{\rm peak} \approx 3.92\,kT$, where $kT$ is the fitted blackbody temperature. For spectra best described by a PL model, $E_{\rm peak}$ falls outside the instrumental bandpass and cannot be directly constrained. In these cases, we examined the spectral evolution using finer temporal slices of approximately equal net photon counts. Based on the evolution of the photon index $\alpha$, we determined whether the power-law component corresponds to the low-energy part of the spectrum below $E_{\rm peak}$, or to the high-energy part above it, and accordingly treated $E_{\rm peak}$ as a lower or upper limit.

\begin{figure*}[htbp]
 \centering
 \plotone{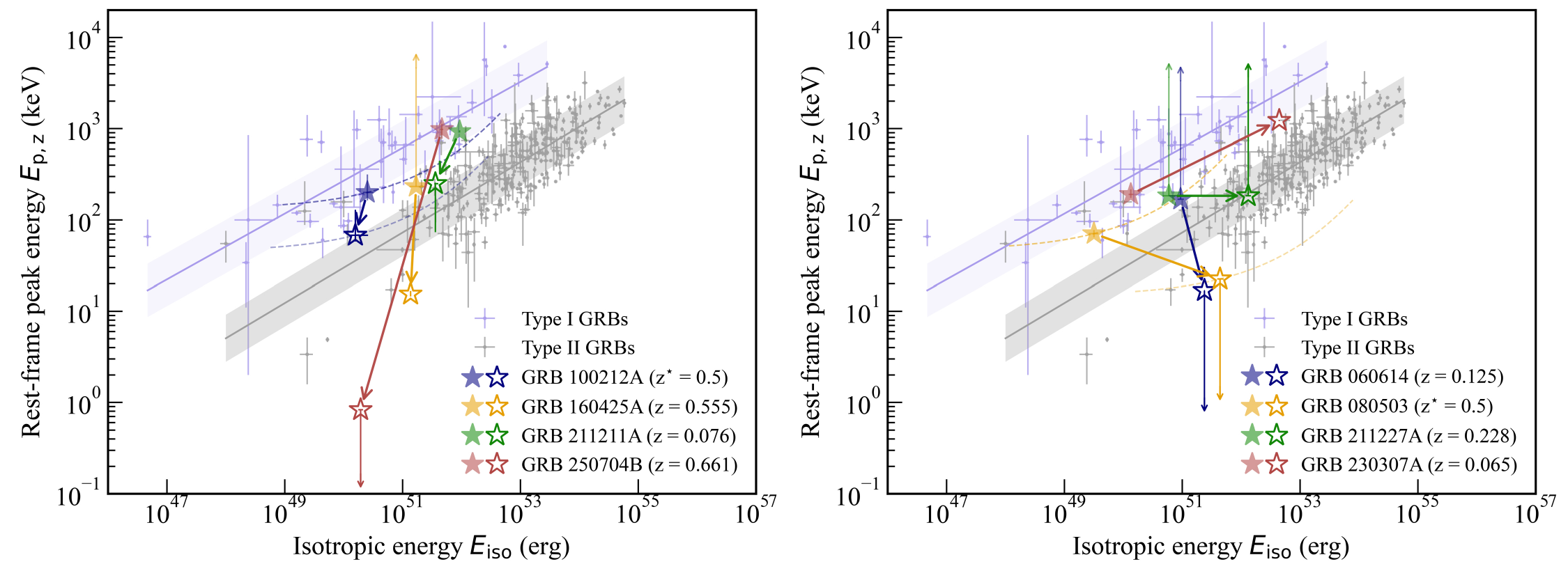}
 \caption{\textbf{Amati-relation diagram for the GRB sample.}
The left panel shows bursts whose transition from Episode~I to Episode~II is accompanied by decreases in both $E_{\mathrm{p,z}}$ and $E_{\mathrm{iso}}$. The right panel shows bursts whose transition toward the Type~II region is accompanied by an increase in $E_{\mathrm{iso}}$. In both panels, filled and open stars denote Episode~I and Episode~II, respectively, and arrows connect the two episodes of the same burst to indicate their episode-dependent transition on the $E_{\rm p,z}$--$E_{\rm iso}$ plane. Vertical arrows denote upper or lower limits on $E_{\mathrm{p,z}}$. For GRB~080503 and GRB~100212A, a reference redshift of $z^{*}=0.5$ is assumed. Dashed lines show the Amati relation at different redshifts, which illustrates the redshift‑dependence for GRB~080503 and GRB~100212A. The blue and gray points, lines, and shaded regions show the empirical Amati-relation distributions for Type~I and Type~II GRBs, respectively.} 
 
\label{fig:amati}
\end{figure*}

Using the preferred spectral model, we computed the bolometric fluence $S_{\rm bol,\gamma}$ over the $1$--$10^{4}$~keV band. The isotropic-equivalent energy $E_{\rm iso}$ was then derived assuming a flat $\Lambda$CDM cosmology with $H_0 = 67.4~\mathrm{km\,s^{-1}\,Mpc^{-1}}$, $\Omega_M = 0.315$, and $\Omega_\Lambda = 0.685$ \citep{2020A&A...641A...6P}. For GRB~080503 and GRB~100212A, which lack redshift measurements, we adopt a fiducial $z=0.5$. Since both $E_\mathrm{peak,z}$ and $E_\mathrm{iso}$ scale with redshift, varying $z$ moves each episode roughly along the Amati relation (dashed lines in Figure~\ref{fig:amati}) rather than across it, so our classification is insensitive to this assumption.

\begin{figure*}[htbp]
\centering
\begin{minipage}[b]{0.31\textwidth}
 \centering
 \includegraphics[width=\linewidth]{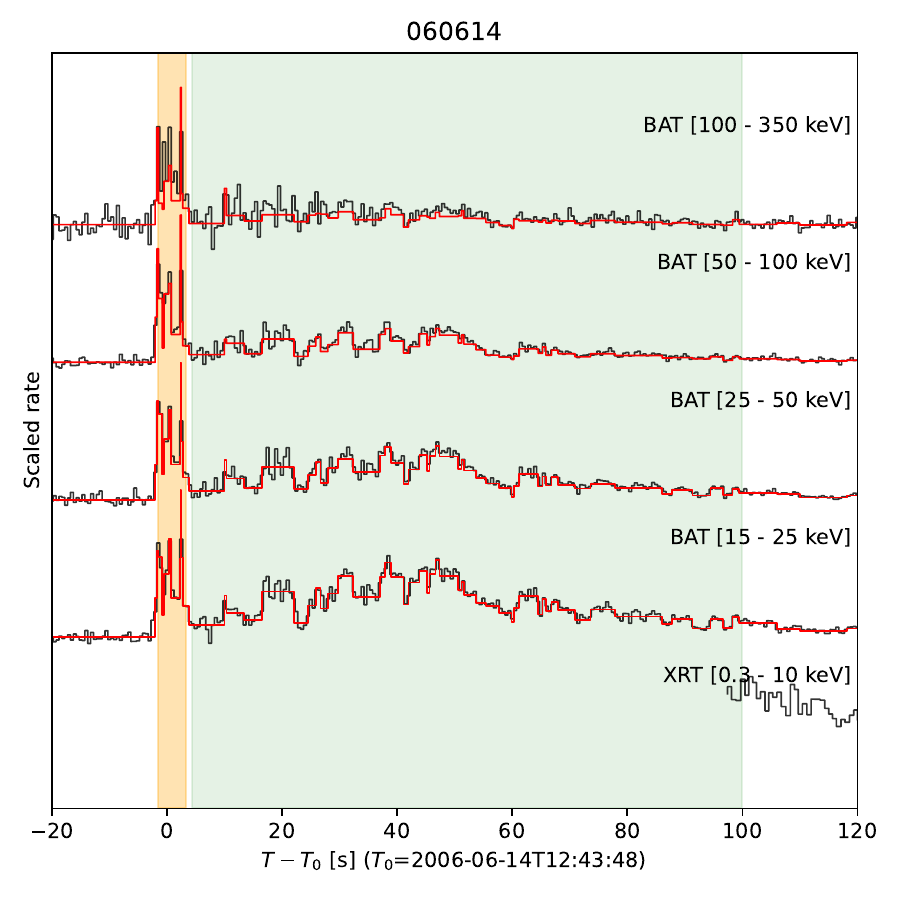}
 \vspace{4pt}
 \includegraphics[width=\linewidth]{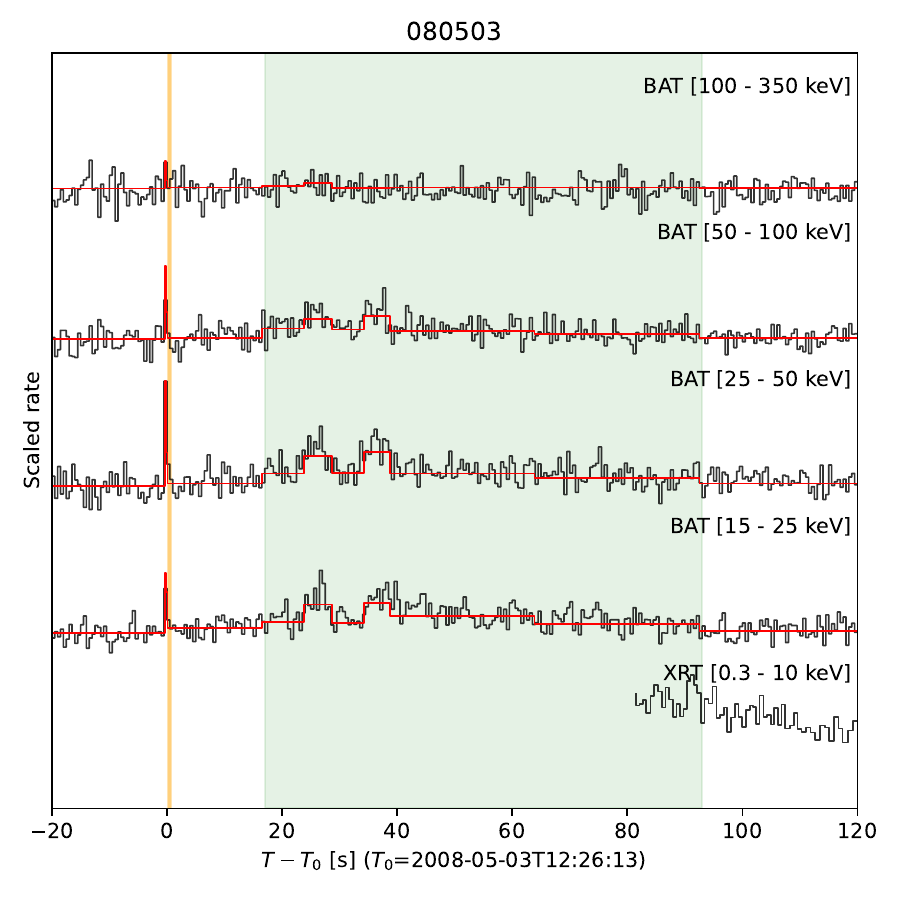}
 \vspace{4pt}
 \includegraphics[width=\linewidth]{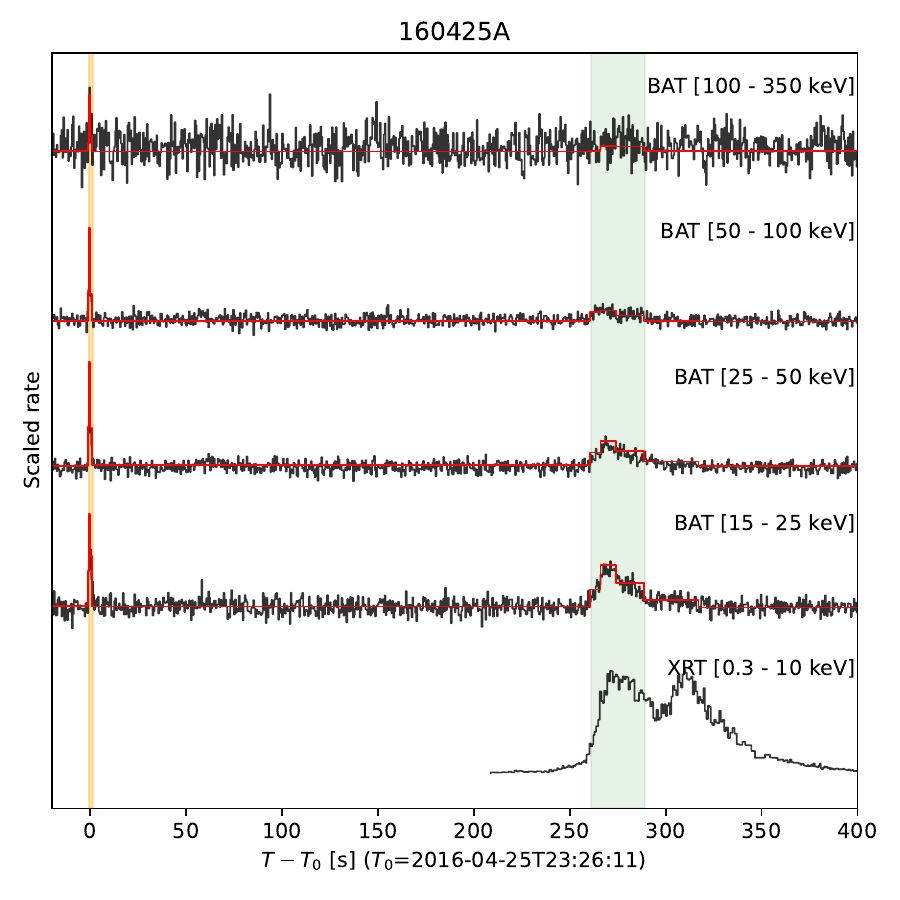}
\end{minipage}\hfill
\begin{minipage}[b]{0.31\textwidth}
 \centering
 \includegraphics[width=\linewidth]{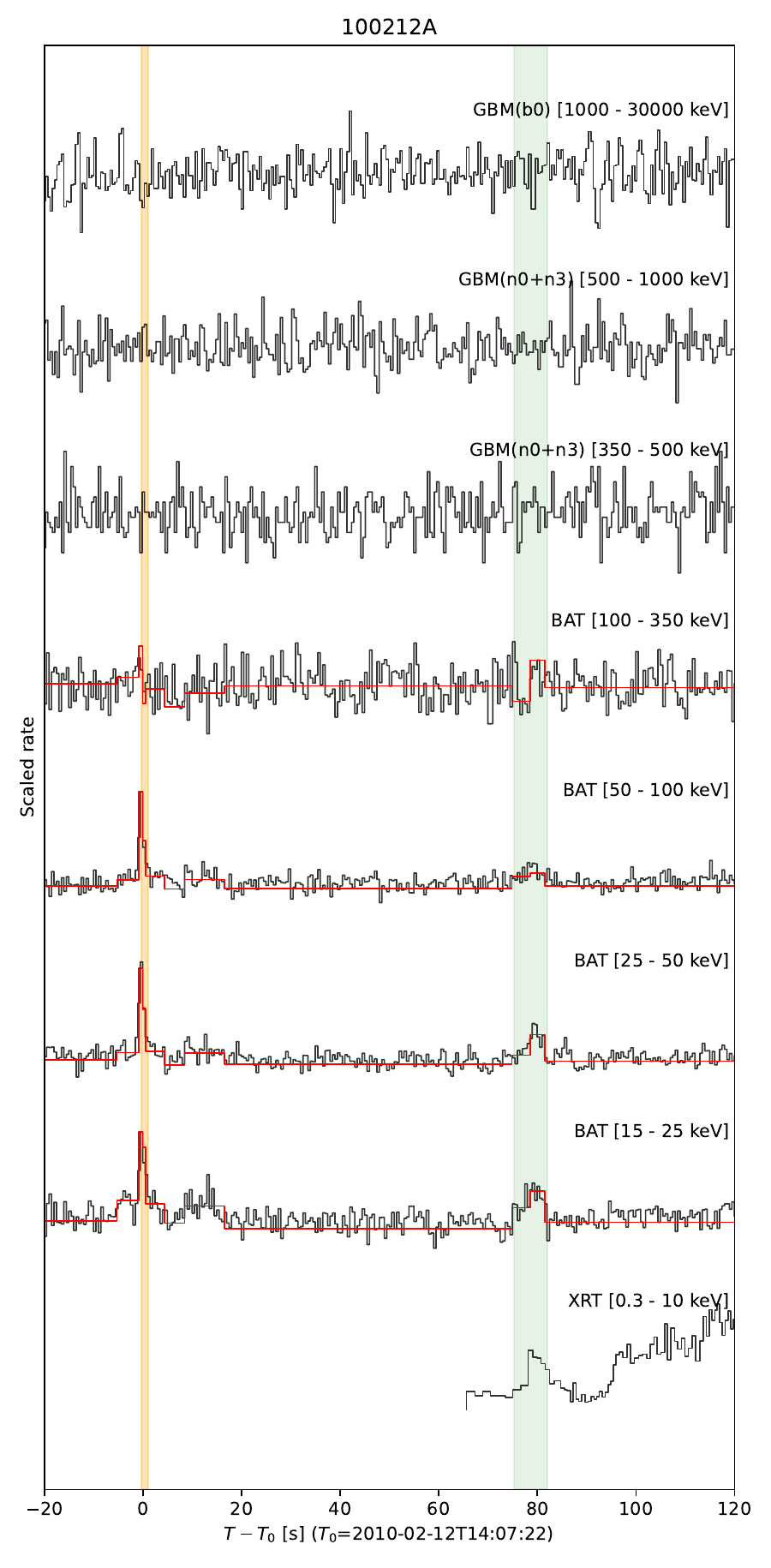}
 \vspace{4pt}
 \includegraphics[width=\linewidth]{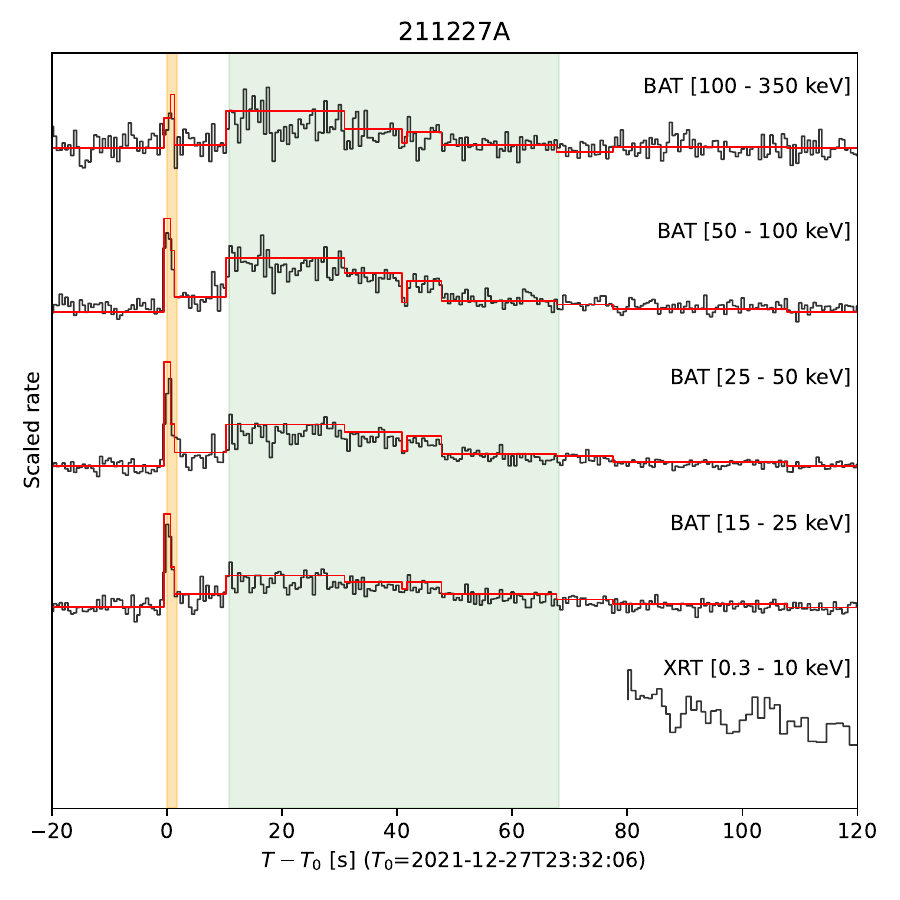}
\end{minipage}\hfill
\begin{minipage}[b]{0.31\textwidth}
 \centering
 \includegraphics[width=\linewidth]{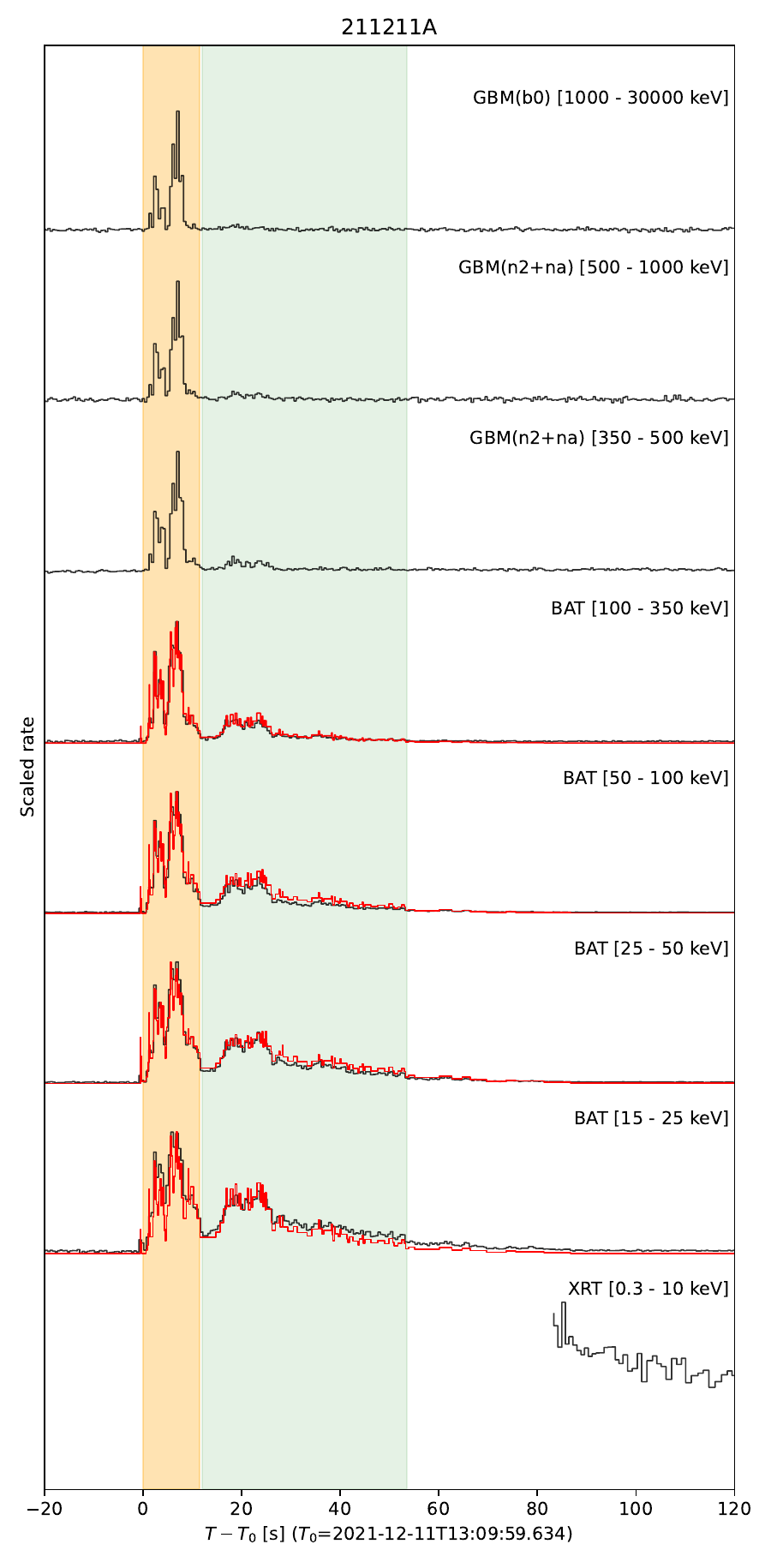}
 \vspace{4pt}
 \includegraphics[width=\linewidth]{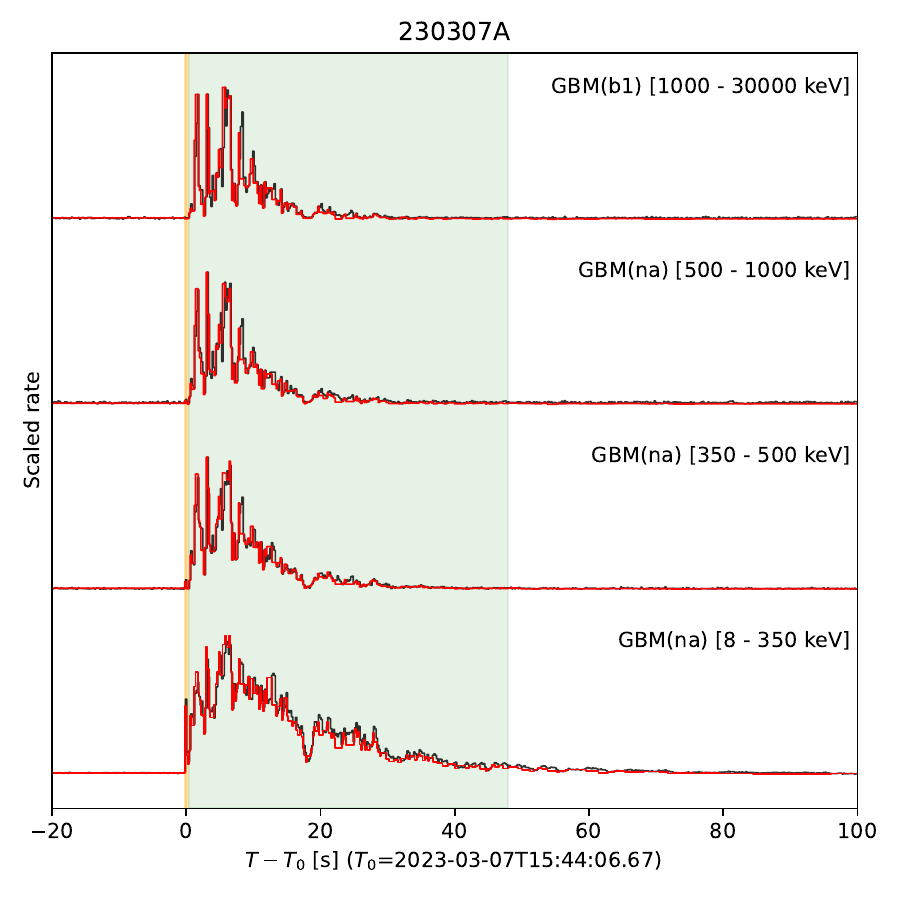}
\end{minipage}
\caption{\textbf{Light curves of the GRB samples adopted in this work.} The black solid lines represent the light curves in different energy bands, while the red solid lines denote the Bayesian Block fittings to BAT data. Notably, GRB 230307A lacks available BAT observations, and its red solid line corresponds to the Bayesian Block results derived from GBM data. The orange shaded regions mark Episode I, and the green shaded regions indicate Episode II. 
Light curve of 250704B can be found in \cite{2026arXiv260114137L}.}
\label{fig:lightcurve}
\end{figure*}

\section{Results} \label{sec:results}

The light curves of the eight selected GRBs, together with the adopted Episode~I and Episode~II intervals, are shown in Figure~\ref{fig:lightcurve}, and the detailed spectral fitting results are provided in Appendix~\ref{spectable}.

Across our sample, Episode~I is generally brief, with durations typically shorter than $\sim$2~s (with the exceptions of GRB~060614 and GRB~211211A, as discussed in Section~\ref{subsec:seg}). This timescale is largely consistent with the canonical definition of short-GRB prompt emission, suggesting that Episode~I resembles the temporal behavior expected from compact-merger events. In contrast, Episode~II lasts for tens to hundreds of seconds --- it is precisely this component that makes these bursts \textit{long-duration} Type~I events --- exhibiting temporal properties more commonly associated with long GRBs.

The multi-band light curves, which for several events span the full energy coverage of BAT, GBM, and XRT, further reveal marked energy-dependent behavior between the two episodes. In most cases, Episode~I is prominent in the high-energy bands, whereas Episode~II is often clearly visible only in the softer, lower-energy channels and fades rapidly toward higher energies. This contrast in energy-band visibility is consistent with a systematic spectral difference between the two components: Episode~I is spectrally hard and dominates at high energies, while Episode~II is substantially softer.

In several bursts, such as GRB~100212A, Episode~I and Episode~II are separated by a relatively quiescent interval. During these intervals, the count rate remains close to the background level before the onset of the later emission. Such gaps can last from several seconds to hundreds of seconds, indicating that the extended emission is not always a smooth continuation of the initial spike. However, a quiescent interval is not required for defining the two episodes. In other events, such as GRB~060614, the transition from the initial spike to the later emission is more continuous. The diversity of temporal morphology therefore suggests that the connection between Episode~I and Episode~II differs from burst to burst, even though the same broad two-component structure is present across the sample.

Building on these temporal and multi-band observations, we now turn to the spectral evolution of the bursts and their behavior on the Amati relation. For each of the eight GRBs, we place Episode~I and Episode~II separately on the $E_{\mathrm{peak,z}}$--$E_{\mathrm{iso}}$ plane, as shown in Figure~\ref{fig:amati}. This allows us to examine whether the empirical classification of a burst depends on which emission episode is used.

Across the sample, Episode~I generally exhibits a hard spectrum and occupies the Type~I region or its vicinity. Episode~II is often softer, especially in events where the extended emission is detected mainly in the X-ray or soft gamma-ray band. However, the spectral evolution is not uniform across the sample: in some events, most notably GRB~230307A, the transition toward the Type~II region is driven primarily by a large increase in $E_{\rm iso}$ rather than by a decrease in $E_{\rm p,z}$.

These episode-dependent spectral and energetic differences lead to different locations on the $E_{\rm p,z}$--$E_{\rm iso}$ plane. Most Episode~I points fall within, or close to, the Type~I region, suggesting a merger-like classification when the initial spike is considered independently. In contrast, most Episode~II points shift toward the Type~II region and would appear Type~II-like if analyzed in isolation.

Taken together, the two episodes reveal an episode-dependent transition across the Amati relation as the emission evolves from the short spike to the later extended component. In several bursts, this transition can be traced directly from the Type~I region toward the Type~II region, linking two empirical regions that are traditionally associated with different progenitor classes. This behavior captures the main observational tension in long-duration Type~I GRBs: the short spike retains the merger-like signature, whereas the later emission can resemble the prompt-emission properties of long GRBs. The amplitude of this transition, however, varies substantially from burst to burst. Some events show a large displacement across the $E_{\rm p,z}$--$E_{\rm iso}$ plane, while others exhibit only modest evolution and remain near the boundary between the two populations.

The detailed transition patterns also differ among individual events. For four bursts in our sample, namely GRBs~100212A, 160425A, 211211A, and 250704B, the transition toward the Type~II region is accompanied by decreases in both $E_{\rm p,z}$ and $E_{\rm iso}$ (left panel of Figure~\ref{fig:amati}). These events therefore move toward the lower-left region of the $E_{\rm p,z}$--$E_{\rm iso}$ plane, broadly consistent with the fading and softening of the burst emission over time.

A different pattern is observed for the remaining four bursts, namely GRBs~060614, 080503, 211227A, and 230307A (right panel of Figure~\ref{fig:amati}). In these events, the transition toward the Type~II region is accompanied by an increase in $E_{\rm iso}$, while $E_{\rm p,z}$ may decrease, remain approximately constant, or increase substantially, as in GRB~230307A. Such evolution therefore cannot be described solely as monotonic spectral softening. Instead, it reflects the joint evolution of spectral hardness and radiation energy, with different bursts following different tracks across the $E_{\rm p,z}$--$E_{\rm iso}$ plane.

We note that several episodes lie near the transition region between the Type~I and Type~II populations and do not clearly belong to either group. Rather than weakening the result, these cases illustrate the continuous nature of the observed transition . Because the Type~I and Type~II populations have substantial intrinsic scatter and partial overlap on the $E_{\rm p,z}$--$E_{\rm iso}$ plane, small differences in episode definition, spectral model, or instrumental energy coverage can shift an event across the empirical boundary. This further highlights the difficulty of assigning a unique progenitor classification based only on the $E_{\rm p,z}$--$E_{\rm iso}$ plane location of a single emission episode.

In addition, several episodes are represented by upper or lower limits on $E_{\mathrm{peak}}$. These limits do not remove the observed transition . In several cases, the allowed values would place Episode~I and Episode~II farther apart on the $E_{\rm p,z}$--$E_{\rm iso}$ plane, suggesting that the measured displacement may be a conservative estimate of the underlying spectral evolution.

A representative example is GRB~211227A, for which both episodes formally yield $E_{\mathrm{peak}}>150$~keV in our BAT-only fitting results. However, the Konus-Wind spectral analysis reported by \citet{2022GCN.31544....1T} indicates that the peak energy of Episode~I is significantly higher than that of Episode~II. Moreover, the spectral and energetic properties of Episode~II remain inconsistent with the typical Type~I region on the $E_{\rm p,z}$--$E_{\rm iso}$ plane. Therefore, although the exact peak energies remain uncertain, the overall transition from a Type~I-like initial spike toward a Type~II-like extended component remains robust.

\section{Discussion} \label{sec:discussion}

\subsection{Synchrotron-Model Constraints on the $E_{\rm p,z}$--$E_{\rm iso}$ transition }

As noted in Section~\ref{sec:data}, the limited energy coverage of BAT can make $E_{\rm peak}$ difficult to constrain for several bursts in our sample. This limitation is especially relevant for Episode~I, whose hard spectrum often places the spectral peak above the BAT bandpass. Empirical spectral fits may therefore provide only upper or lower limits on $E_{\rm peak}$, introducing uncertainty in the inferred location of the episode on the $E_{\rm p,z}$--$E_{\rm iso}$ plane. To test whether the episode-dependent transition identified in Section~\ref{sec:results} persists under a more physically constrained treatment of the spectral evolution, we performed an additional physical-model analysis of GRB~160425A as a case study. GRB~160425A is selected for two reasons. Spectrally, only BAT data are available for this burst, so that empirical fits provide only loose constraints on $E_\mathrm{peak}$, making it a natural test case for the physically motivated synchrotron model. Temporally, both of its emission episodes show relatively simple, FRED-like pulse morphologies, as required by the synchrotron model adopted here, which allows a consistent time-resolved physical modelling across the two episodes.

For this purpose, we adopted the synchrotron emission framework developed by \citet{2023ApJ...947L..11Y,2024ApJ...962...85Y}. Unlike purely empirical spectral fitting, this model fits the coupled temporal and spectral evolution of an emission pulse, thereby providing additional constraints on the evolution of the characteristic spectral energy. This is particularly useful when the instantaneous empirical spectrum is only weakly constrained by the instrumental bandpass. For GRB~160425A, we modeled the single-pulse structure of Episode~I and the subsequent extended-emission phase using time-resolved spectra, and derived the corresponding evolution of $E_{\rm peak}$ and $E_{\rm iso}$ (Figures~\ref{fig:160425A}, \ref{fig:syn_g1}, and \ref{fig:syn_g2}). The synchrotron fitting details are presented in Appendix~\ref{synchrotron}.

The resulting time-resolved $E_{\rm p,z}$--$E_{\rm iso}$ plane evolution is shown in Figure~\ref{fig:160425A}. The synchrotron-model constraints are broadly consistent with the empirical Band-function results presented in Section~\ref{sec:results}: the burst evolves from a Type~I-like location during Episode~I toward a Type~II-like location during Episode~II. More specifically, the model traces a progressive shift toward lower $E_{\mathrm{peak}}$ and larger $E_{\mathrm{iso}}$, reproducing the same overall transition inferred from the empirical analysis. This behavior is seen not only between the two major emission episodes, but also within the finer time-resolved evolution of the burst.

Our inferred $E_{\rm p}$ for Episode~II ($\sim$10~keV) is lower than the value reported by \citet{2026arXiv260328699L} ($\sim$ 57.6~keV). A plausible reason is that our analysis includes XRT data, which extend the low-energy coverage and provide stronger constraints on the soft extended-emission component. This difference highlights the same observational issue emphasized throughout this work: the inferred position of a burst on the $E_{\rm p,z}$--$E_{\rm iso}$ plane can depend sensitively on the available energy coverage and on which emission component is included.

The synchrotron modeling does not by itself determine the progenitor type of GRB~160425A. Rather, it provides an independent check that the observed $E_{\rm p,z}$--$E_{\rm iso}$ transition is not simply an artifact of poorly constrained empirical $E_{\mathrm{peak}}$ values. This point is consistent with the broader classification framework advocated by \citet{2009ApJ...703.1696Z}, in which Type~I/II classification should be based on multiple observational criteria rather than on duration, hardness, or a single prompt-emission correlation alone. In this context, the synchrotron-model result strengthens our main conclusion: for long-duration Type~I GRBs, the short spike and the later extended emission can occupy different regions of the $E_{\rm p,z}$--$E_{\rm iso}$ plane, and the apparent classification can change depending on which part of the burst is observed.

\begin{figure*}[ht!]
\plotone{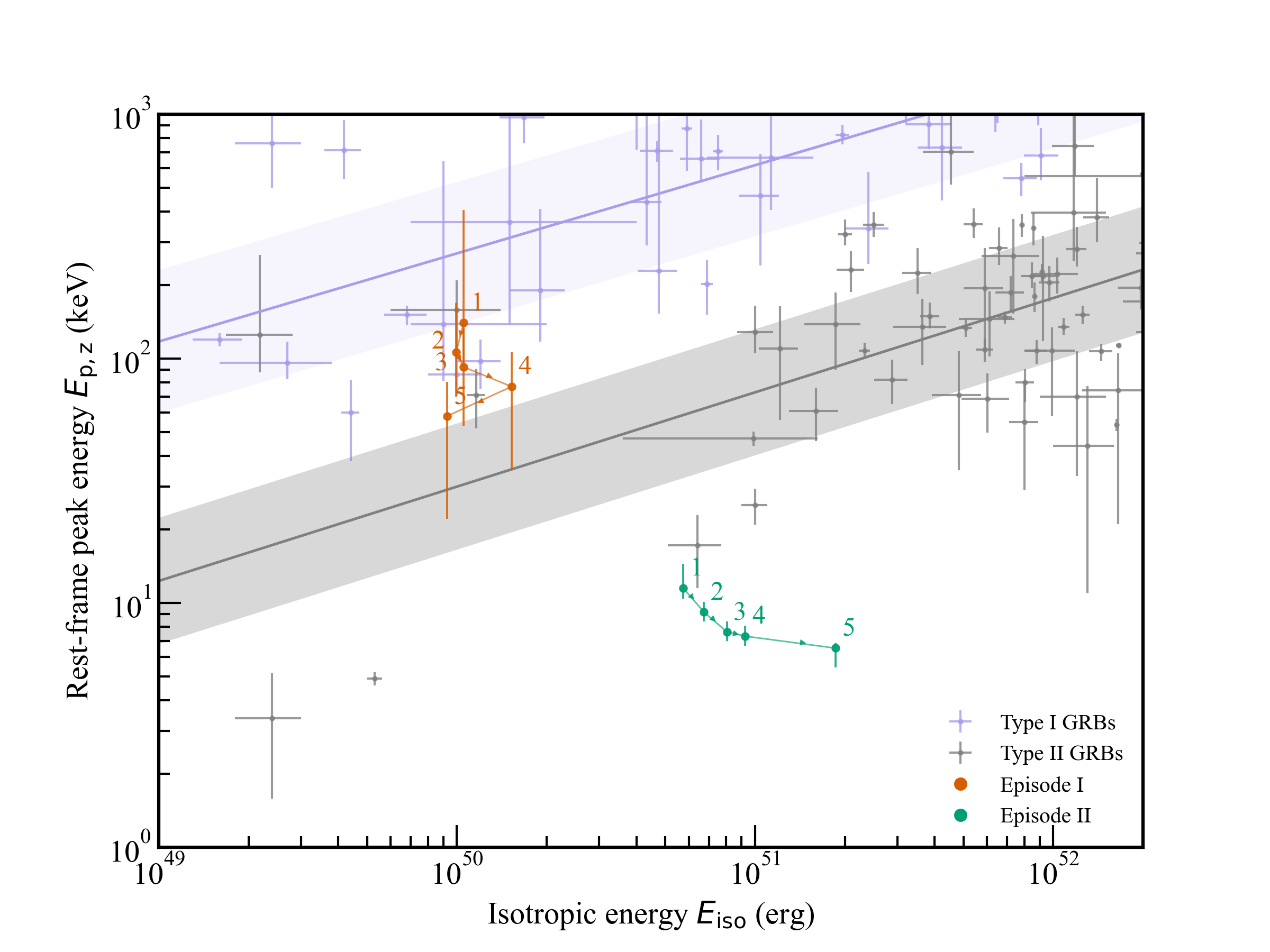}
\caption{\textbf{Time-resolved $E_{\rm p,z}$--$E_{\rm iso}$ plane evolution of GRB~160425A derived from synchrotron modeling.}
Orange and green points mark successive time slices of Episode~I and Episode~II, respectively. Arrows indicate the temporal evolution. The synchrotron-model constraints show a progressive transition from a Type~I region during the initial spike toward a Type~II region during the extended-emission phase.}
\label{fig:160425A}

\end{figure*}

\subsection{Misclassification Risks and Observational Biases}

The episode-dependent transition identified in Section~\ref{sec:results} has a practical consequence: the Amati-based classification of a long-duration Type~I GRB can change substantially depending on which emission episode is detected. This introduces a systematic misclassification risk that operates in both directions.

One important risk is that the merger origin of a long-duration Type~I GRB may be missed entirely. If the initial short hard spike --- the episode that carries the Type~I signature --- is weak, falls outside the instrumental bandpass, or does not trigger the detector, the observed emission will be dominated by the softer extended component. In that case, both the duration and the $E_{\rm p,z}$--$E_{\rm iso}$ plane location of the burst will point toward a collapsar origin. GRB~250704B illustrates this scenario directly: as discussed in \citet{2026arXiv260114137L}, this merger-related burst exhibits several hundred seconds of extended soft X-ray emission detected by EP/WXT, whose time-integrated spectral properties place it within the Type~II region. Had the initial hard gamma-ray spike not been detected --- and without independent multi-wavelength diagnostics --- the burst would have been classified as a canonical long GRB despite its likely merger origin.

The reverse misclassification is also possible, though not directly illustrated by our sample. If the observed emission of a long-duration Type-II GRB is dominated by an initial hard episode while the softer extended emission remains below the instrumental sensitivity, the burst may appear short in duration and occupy the Type~I region of the $E_{\rm p,z}$--$E_{\rm iso}$ plane, potentially mimicking a compact-merger event. 

These considerations reinforce a point already raised in the context of individual bursts such as GRB~250704B \citep{2026arXiv260114137L} and GRB~160425A \citep{2026arXiv260328699L}: neither burst duration nor the Amati relation alone is a reliable progenitor diagnostic for long-duration Type~I GRBs. Broad energy coverage spanning from the hard gamma-ray to the soft X-ray band, high temporal resolution to resolve individual emission episodes, and independent multi-wavelength diagnostics such as kilonova searches and host-galaxy characterization are all necessary to robustly identify the physical origin of these complex events.

\subsection{Possible Physical Origins of the Episode-Dependent Transition}

The episode-dependent transition indicates that the location of a burst on the $E_{\rm p,z}$--$E_{\rm iso}$ plane is strongly affected by the spectral properties of the emission episode selected for analysis. It should therefore not be interpreted as a fixed property of the progenitor system.

For relatively simple cases, the observed transition may be explained by the luminosity decline and spectral softening of a single relativistic outflow as it evolves over time \citep{2012ApJ...757...56Y, 2018ApJ...869..100U}. In this picture, Episode~I and Episode~II represent early and late phases of the same jet, with the $E_{\rm p,z}$--$E_{\rm iso}$ plane transition reflecting the continuous fading and cooling of the emission.

However, several events in our sample are difficult to account for by this simple picture alone. In particular, the well-defined quiescent interval followed by renewed pulsed emission in GRB~160425A \citep{2026arXiv260328699L} more naturally suggests intermittent central-engine activity \citep{2001MNRAS.324.1147R} rather than continuous evolution of a single outflow component. For such events, additional physical ingredients may be required, including variable late-time accretion onto the central remnant \citep{2009ApJ...696.1871P}, changes in jet magnetization or baryon loading between episodes \citep{2011MNRAS.413.2031M}, or transitions between distinct radiation mechanisms \citep{2015ApJ...807..163G, 2025NSRev..12E.401S}. Distinguishing among these scenarios will require broadband, episode-resolved spectroscopy of a larger sample of similar events.

\section{Summary and Conclusion} \label{sec:conclusion}

In this work, we systematically investigated eight long-duration Type-I GRBs and merger candidates, each exhibiting an initial short hard spike (Episode~I) followed by softer extended emission (Episode~II). By separating the two episodes and tracking their individual locations on the $E_{\rm p,z}$--$E_{\rm iso}$ plane, we tested whether the widely used Amati relation remains a reliable progenitor diagnostic when a burst's emission history is more complicated than a single episode. Our main findings are as follows:

\begin{itemize}
\renewcommand{\labelitemi}{-}

\item The short spike tells only part of the story. All eight bursts display two distinct emission components with systematically different spectral properties: a brief, spectrally hard Episode~I that falls in or near the Type~I region of the $E_{\rm p}$--$E_{\rm iso}$ plane, and a longer-lasting, softer Episode~II that shifts toward the Type~II region. If analyzed in isolation, the two episodes of the same burst would lead to opposite empirical classifications---yet a compact-merger and a collapsar cannot both be the progenitor of a single event. This contradiction exposes a fundamental limitation of the Amati relation as a standalone progenitor diagnostic for long-duration Type~I GRBs.

\item The Type~I-to-Type~II transition does not follow a single track. In four bursts, GRBs~100212A, 160425A, 211211A, and 250704B, both $E_{\rm p,z}$ and $E_{\rm iso}$ decrease from Episode~I to Episode~II, consistent with fading and softening. In the other four bursts, GRBs~060614, 080503, 211227A, and 230307A, the transition is accompanied by an increase in $E_{\rm iso}$, indicating that the evolution involves both spectral and energetic changes.

\item The observed transition is unlikely to be solely an artifact of poorly constrained $E_{\rm peak}$ values. For GRB~160425A, synchrotron-model fitting reproduces the same progressive Type~I-to-Type~II evolution inferred from the empirical analysis. In addition, several $E_{\rm peak}$ limits would place the two episodes farther apart on the $E_{\rm p,z}$--$E_{\rm iso}$ plane, making the measured displacement a conservative estimate.

\item Amati relation-based classification is therefore subject to an episode-selection bias. The inferred position of a burst on the $E_{\rm p,z}$--$E_{\rm iso}$ plane depends strongly on which emission episode is detected. If the short hard spike is weak, falls outside the instrumental bandpass, or is missed entirely --- as could easily have occurred for GRB~250704B without its gamma-ray trigger --- a merger-origin event would be classified as a canonical long GRB-like event based on its softer extended emission alone.

\end{itemize}

This work is subject to several limitations: the sample remains small, and the limited instrumental energy coverage introduces uncertainties in the spectral characterization of some episodes. Larger samples --- particularly those enabled by wide-field soft X-ray monitors such as the Einstein Probe --- will be essential for establishing how common the episode-dependent transition is and for clarifying its physical origin. Nevertheless, our results offer a clear caution: for long-duration Type~I GRBs, neither the burst duration nor the Amati relation alone should be interpreted as a direct progenitor diagnostic. Reliable classification requires broad temporal and spectral coverage that captures the full emission history together with independent multi-wavelength diagnostics.

\begin{acknowledgments}
We acknowledge the support by the National Natural Science Foundation of China (grants 12573046 and 12121003 to Bin-Bin Zhang and grants 12393811 and 12473048 to Xiao-Hong Zhao), the science research grants from the China Manned Space Project (grant CMS-CSST-2021-B11 to Bin-Bin Zhang). Bin-Bin Zhang acknowledges support by the Fundamental Research Funds for the Central Universities, and the Programme for Innovative Talents and Entrepreneurs in Jiangsu. Zhen-Yu Yan is supported by the Program for Outstanding PhD Candidates of Nanjing University (2025A06).
\end{acknowledgments}

\appendix

\section{Synchrotron fitting results of GRB 160425A} \label{synchrotron}

\begin{figure*}[ht!]
\plotone{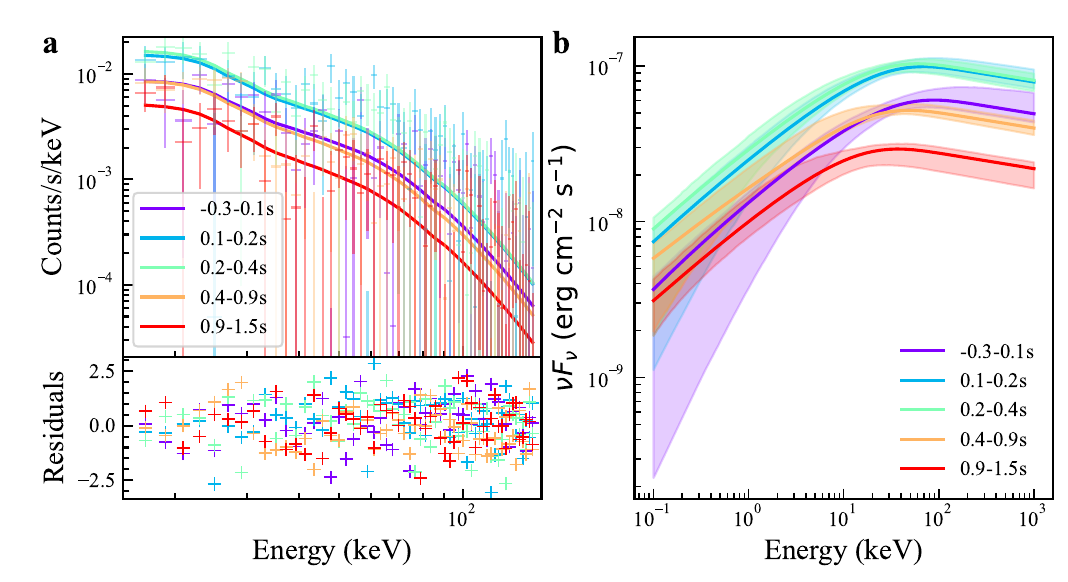}
\caption{\textbf{Synchrotron modeling of GRB~160425A Episode~I.}
\textbf{a.} Observed photon count spectrum and the best-fit synchrotron model prediction.
\textbf{b.} Time-resolved model $\nu F_{\nu}$ spectra derived from the synchrotron fit.
xError bars and shaded regions denote $1\sigma$ uncertainties.}
\label{fig:syn_g1}

\end{figure*}

\begin{figure*}[ht!]
\plotone{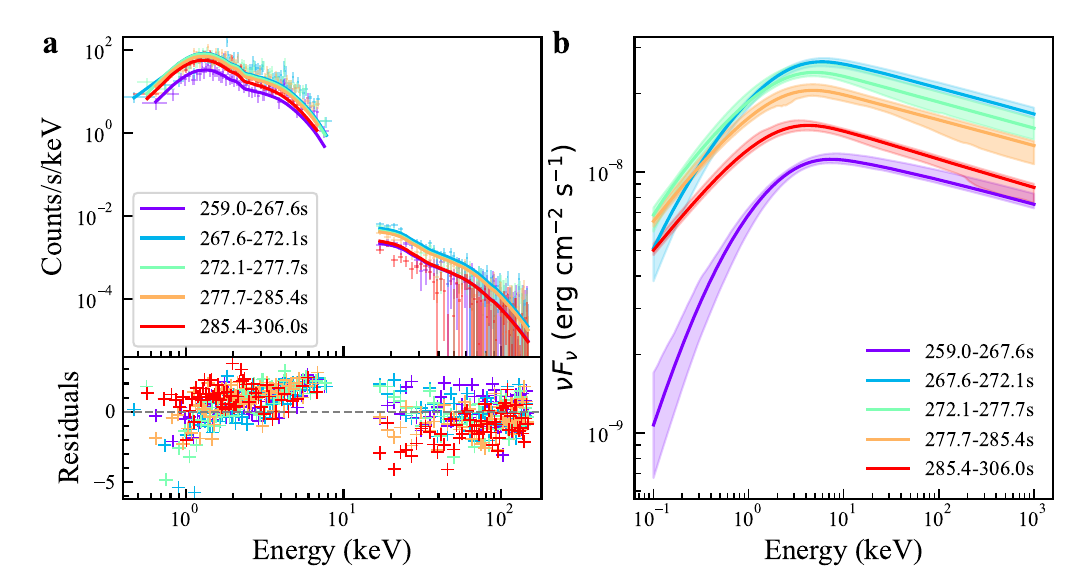}
\caption{\textbf{Synchrotron modeling of GRB~160425A Episode~II.}
\textbf{a.} Observed photon count spectrum and the best-fit synchrotron model prediction, including the XRT data that extend the spectral coverage to lower energies.
\textbf{b.} Time-resolved model $\nu F_{\nu}$ spectra derived from the synchrotron fit.
Error bars and shaded regions denote $1\sigma$ uncertainties.}
\label{fig:syn_g2}

\end{figure*}

A time-dependent synchrotron model was used to perform a physical fit for GRB 160425A. To mitigate parameter degeneracy, we fixed two parameters in the framework developed by \citet{2023ApJ...947L..11Y,2024ApJ...962...85Y}, $\alpha_B=1.0$ and $p=2.2$. The spectra and light curves for Episode I and II are shown in Figure \ref{fig:syn_g1} and \ref{fig:syn_g2}, respectively.

\clearpage

\section{Spectral Fitting Results} \label{spectable}

\begin{sidewaystable}[h]
\centering
\caption{Spectral fitting results}
\begin{tabular}{ccc cccc cccc}
\hline
\hline

 & & & \multicolumn{4}{c}{Model Comparison (BIC)} & \multicolumn{4}{c}{Best Model Parameter} \\
\cline{4-7} \cline{8-11}

 GRB name & Duration & Band & PL & CPL & BB & BAND & $\alpha$ & $\beta$ & $E_{\mathrm{peak}}$ (keV) & $S_{\mathrm{bol,\gamma}}$ (erg $\mathrm{cm}^{-2}$) \\

\hline

060614 & [-1.55, 3.25] & BAT & 61.086 & 65.154 & 273.250 & & $-1.618^{+0.04}_{-0.04}$ & & $>150$ & $2.41^{+0.33}_{-0.28} \times10^{-5}$ \\
& [4.34, 99.95] & BAT & 57.106 & 65.642 & 996.25 & & $-2.071^{+0.023}_{-0.023}$ & & $<15$ & $6.12^{+0.07}_{-0.07} \times10^{-5}$ \\
080503 & [0.11, 0.62] & BAT & 33.484 & 31.859 & 29.948 & & & & $47.129^{+4.547}_{-4.048}$ $^{\textsuperscript{a}}$ & $4.66^{+0.91}_{-0.86} \times10^{-8}$ \\
& [17.07, 93.04] & BAT & 48.119 & 52.053 & 142.915 & & $-1.9^{+0.075}_{-0.064}$ & & $<15$ & $6.44^{+0.95}_{-0.66} \times10^{-6}$ \\
100212A & [-0.38, 1.02] & BAT+GBM & 119.861 & 107.611 & 163.594 & 112.048 & $-1.069^{+0.163}_{-0.235}$ & & $133.626^{+76.977}_{-24.204}$ & $3.71^{+0.97}_{-0.57} \times10^{-7}$ \\
& [75.33, 81.98] & BAT+GBM & 77.061 & 74.958 & 79.229 & 78.231 & $-0.818^{+0.444}_{-0.936}$ & & $45.273^{+40.976}_{-6.638}$ & $2.32^{+1.49}_{-0.53} \times10^{-7}$ \\
160425A & [-0.544, 1.376] & BAT & 34.069 & 37.846 & 86.108 & & $-1.672^{+0.126}_{-0.083}$ & & $>150$ & $2.01^{+0.91}_{-0.46} \times10^{-6}$ \\
& [260.96, 289.12] & BAT+XRT & 642.484 & 579.492 & & 465.762 & $-0.851^{+0.171}_{-0.126}$ & $-2.561^{+0.042}_{-0.095}$ & $9.931^{+0.834}_{-0.555}$ & $1.62^{+0.05}_{-0.04} \times10^{-6}$ \\
211211A & [0.03, 11.49] & BAT+GBM & 35968.093 & 3271.398 & 99982.941 & 2672.175 & $-1.007^{+0.006}_{-0.007}$ & $-2.459^{+0.031}_{-0.053}$ & $861.608^{+28.834}_{-21.651}$ & $6.59^{+0.05}_{-0.13} \times10^{-4}$ \\
& [12.06, 53.47] & BAT+GBM & 11215.908 & 9526.369 & 69051.571 & 9509.345 & $-1.619^{+0.515}_{-0.006}$ & $-2.452^{+0.442}_{-0.06}$ & $232.613^{+6.132}_{-164.442}$ & $2.56^{+0.31}_{-0.96} \times10^{-4}$ \\
211227A & [0.03, 1.82] & BAT & 58.230 & 61.109 & 100.572 & & $-1.63^{+0.089}_{-0.075}$ & & $>150$ & $4.43^{+1.42}_{-0.95} \times10^{-6}$ \\
& [10.72, 68.19] & BAT & 59.394 & 61.07 & 347.588 & & $-1.426^{+0.029}_{-0.03}$ & & $>150$ & $9.84^{+1.13}_{-0.97} \times10^{-5}$ \\
230307A & [-0.2, 0.2] & GECAM-B & 285.892 & 235.689 & 416.031 & 241.295 & $-1.107^{+0.169}_{-0.199}$ & & $179.094^{+24.112}_{-14.819}$ & $1.31^{+0.15}_{-0.11} \times10^{-5}$ \\
& [0.4, 48] & GECAM-B & 8339.020 & 646.472 & 42521.497 & 652.595 & $-1.154^{+0.006}_{-0.005}$ & & $1151.081^{+18.27}_{-17.455}$ & $4.32^{+0.04}_{-0.04} \times10^{-3}$ \\
250704B$^{\textsuperscript{b}}$ & [-0.10, 0.40] & SVOM/GRM & & & & & $-0.78^{+0.31}_{-0.27}$ & $-1.46^{+0.22}_{-0.21}$ & $588.84^{+91.93}_{-122.18}$ & $3.15^{+0.17}_{-0.21} \times10^{-6}$ \\
& [22.76, 764.50] & EP/WXT & & & & & $-2.30^{+0.23}_{-0.23}$ & & $<0.5$ & $1.61^{+0.16}_{-0.14} \times10^{-7}$ \\

\hline
\hline
\multicolumn{11}{c}{\footnotesize $^{\textsuperscript{a}}$ The peak energy $E_{\mathrm{peak}}$ here is calculated through parameter \textit{temperature} of Black Body model.} \\
\multicolumn{11}{c}{\footnotesize $^{\textsuperscript{b}}$ Refer to \cite{2026arXiv260114137L}.} \\

\end{tabular}

\end{sidewaystable}

\bibliography{sample701}{}
\bibliographystyle{aasjournalv7}

\end{document}